\documentclass{svproc}
\usepackage{url}
\usepackage{graphicx}
\usepackage{subfigure}
\usepackage{booktabs}
\usepackage{multirow}
\usepackage{multicol}
\usepackage{lipsum}
\usepackage{color}
\usepackage{hyperref}

\begin{document}
\mainmatter              
%
\title{3D Displays: Their Evolution, Inherent Challenges \& Future Perspectives}
%

\author{Xingyu Pan \and Xuanhui Xu \and Soumyabrata Dev \and Abraham G Campbell}

\authorrunning{Pan et al} 

\institute{School of Computer Science\\University College Dublin, Ireland \\
\email{xingyu.pan@ucdconnect.ie},
\email{xuanhui.xu@ucdconnect.ie},
\email{soumyabrata.dev@ucd.ie},
\email{abey.campbell@ucd.ie}
}

\maketitle              

\begin{abstract}
The popularity of 3D displays has risen drastically over the past few decades but these displays are still merely a novelty compared to their true potential. The development has mostly focused on Head Mounted Displays (HMD) development for Virtual Reality and in general ignored non-HMD 3D displays. This is due to the inherent difficulty in the creation of these displays and their impracticability in general use due to cost, performance, and lack of meaningful use cases. In fairness to the hardware manufacturers who have made striking innovations in this field, there has been a dereliction of duty of software developers and researchers in terms of developing software to best utilize these displays. 

This paper will seek to identify what areas of future software development could mitigate this dereliction.
To achieve this goal, the paper will first examine the current state of the art and perform a comparative analysis on different types of 3D displays, from this analysis a clear researcher gap exists in terms of software development for Light field displays which are the current state of the art of non-HMD-based 3D displays.

The paper will then outline six distinct areas where the context-awareness concept will allow for non-HMD-based 3D displays in particular light field displays that can not only compete but surpass their HMD-based brethren for many specific use cases.

\end{abstract}

\section{Introduction}
In recent years, 3D displays are now playing an increasing role in 3D visualization, education, and entertainment. Virtual and Augmented Reality has entered the mainstream with Facebook's Oculus Rift and Microsoft Hololens acting as the poster children for this new revolution in Human Computer Interaction (HCI). These technologies through requiring a user to wear a device to provide solitary experiences unless the software is written to allow for group experiences. They produce a facsimile of the mythic Holodeck experiences that were promised by sciences fiction authors in the past, but it is only a solitary facsimile, and as long as there is a device a user needs to wear it can never be the ultimate display~\cite{sutherland1965ultimate}. 

Even with the success of HMD-based 3D displays, work has continued on non-HMD-based 3D display technologies, with considerable success in terms of hardware development of auto-stereoscopic displays. The advancement of what is called Super-Multi-View (SMV) auto-stereoscopic displays, can now approximately output a sample of a light field. It is only a sample due to the true analogy nature of a light field which can be considered a continuous function that records all light entering a single point. This function consists of five dimensions and was proposed by Andreī Gershun~\cite{gershun1939light} in 1939. This can be sampled as shown by the work of Levoy and Hanrahan~\cite{levoy1996light} where they demonstrate light field rendering is possible. Since Super-Multi-View (SMV) auto-stereoscopic displays are capable of approximating a light field, the term Light Field Displays (LFDs) has begun to be used, not only by the companies themselves manufacturing the displays but by researchers in the field~\cite{urey2011state}. This can be considered controversial though as mentioned already that LFDs are only approximating a light field so the distinction of the "level of sampling" necessary to be considered a true light field display becomes one of academic debate.  
 
These new Light Field Displays, best exemplified by the Looking Glass display have to overcome some major intrinsic challenges before mainstream adoption can take place. The hardware has matured but the software has not. One can argue that there has been a major dereliction of duty of software developers and researchers in terms of developing software to best utilizing these displays. This paper hopes to contribute to this field in several ways. Its first contribution is to give a background review of current 3D displays, then performing comparative analysis on these displays. Its final contribution is to use this analysis to generate possible future directions for LFD software development in particular because this is an area of research that has been neglected. With more research, this paper proposes that the future of 3D displays is only just beginning and the LFDs will not only compete with their fellow 3D display HMD-based brethren but in certain use cases surpass them.

The cornerstone of this contribution is applying the concept of context-awareness of the user or users in respect to the display. In other 3D displays, this context-awareness is simple a byproduct necessary to allow the displays to work, but with LFDs it is not naturally present. With its addition, six future directions of research can be identified. 

The remainder of this paper is structured as follows. In Section~\ref{sec:Background}, the overall background of 3D displays will be discussed, which includes the development and evolution of Stereoscopic 3D displays that assist the viewers with binocular parallax. Based on this background, this Section delves further into the usage of non-stereoscopic 3D displays and introduces formally the next-generation autostereoscopic displays and light field displays. In Section~\ref{sec:discuss} a comparative analysis of the various 3D displays will be performed. With this analysis completed, the six potential future directions for LFDs will be outlined in Section~\ref{sec:future}. Finally, Section~\ref{sec:conc} concludes the paper. 

\section{Background on 3D displays}
\label{sec:Background}
Although the conventional 2D displays still occupy the mainstream, however over the past decades, different types of 3D displays appeared and gained an increasing trend to replace the conventional 3D displays~\cite{urey2011state}. There are several factors that assist the human visual system to perceive 3D scenes. Nick Holliman in \cite{holliman20053d} has summarized these factors/cues that assist in 3D displays. Primarily, these cues include: 
\begin{itemize}
    \item Monocular depth cues: Monocular depth cues are the most common depth cues that the conventional 2D displays provide to the viewers. These cues indicate the depth perception to the viewers based on their understanding of the world. The displays render 3D contents with the similar perspective relationship as the real world. This type of depth cues works well on displaying the real-world scenes, however, the viewers are hard to reconstruct the 3D contents in their brains when complicated scientific models are displayed. 
    \item Oculomotor depth cues:
    Oculomotor depth cues consist of accommodation, convergence, and myosis~\cite{reichelt2010depth}. When looking at objects at a different depth, the muscles of the human eyes will adjust these cues to form clear images on the retina. As a result, human brains can infer the depth of information according to these cues.
    \item Binocular parallax: Binocular parallax is also known as binocular disparity. When looking at a 3D scene, each of the eyes will see similar but slightly different scenes. The brains then automatically analyze the difference between the two images and output the depth information.
    \item Motion parallax: Motion parallax indicates the objects are in 3D when the viewer moves his/her head. Viewers will see a different aspect of the objects at a different position, therefore, their brains are able to reconstruct the 3D object from these continuous images.
    
\end{itemize} 
The conventional 2D displays can only provide the viewers with the monocular depth cues. However, the 3D displays can always provide at least one more type of depth cues. With a context of cues established now a discussion on the evolution of Stereoscopic 3D displays can take place. 

\subsection{Evolution of Stereoscopic 3D displays for binocular parallax}
\label{sec:dev}
    Stereoscopic 3D displays provide the viewers with binocular parallax. According to the light source position, there are two main types of stereoscopic 3D displays: head-mounted-displays (HMDs) and glasses-required displays (GRDs).  
    \subsubsection{Head-Mounted Displays (HMDs):} 
\begin{figure*}[ht]
    \subfigure[Sega VR (1995)]{
    \begin{minipage}[htbp]{0.3\linewidth}
    \includegraphics[height=3cm]{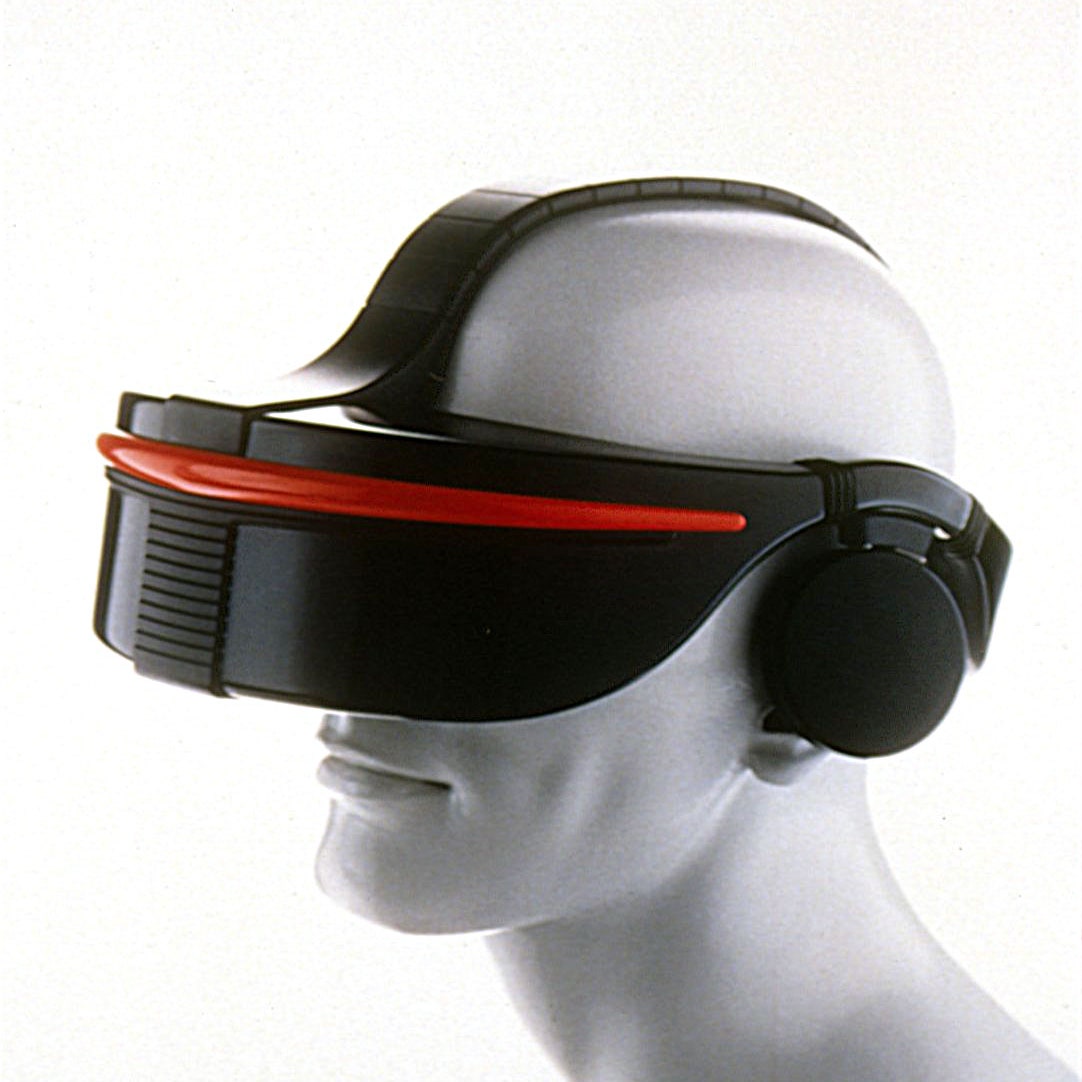}
    \end{minipage}
    }
    \subfigure[Virtual Boy (1995)]{
    \begin{minipage}[htbp]{0.33\linewidth}
    \includegraphics[height=3cm]{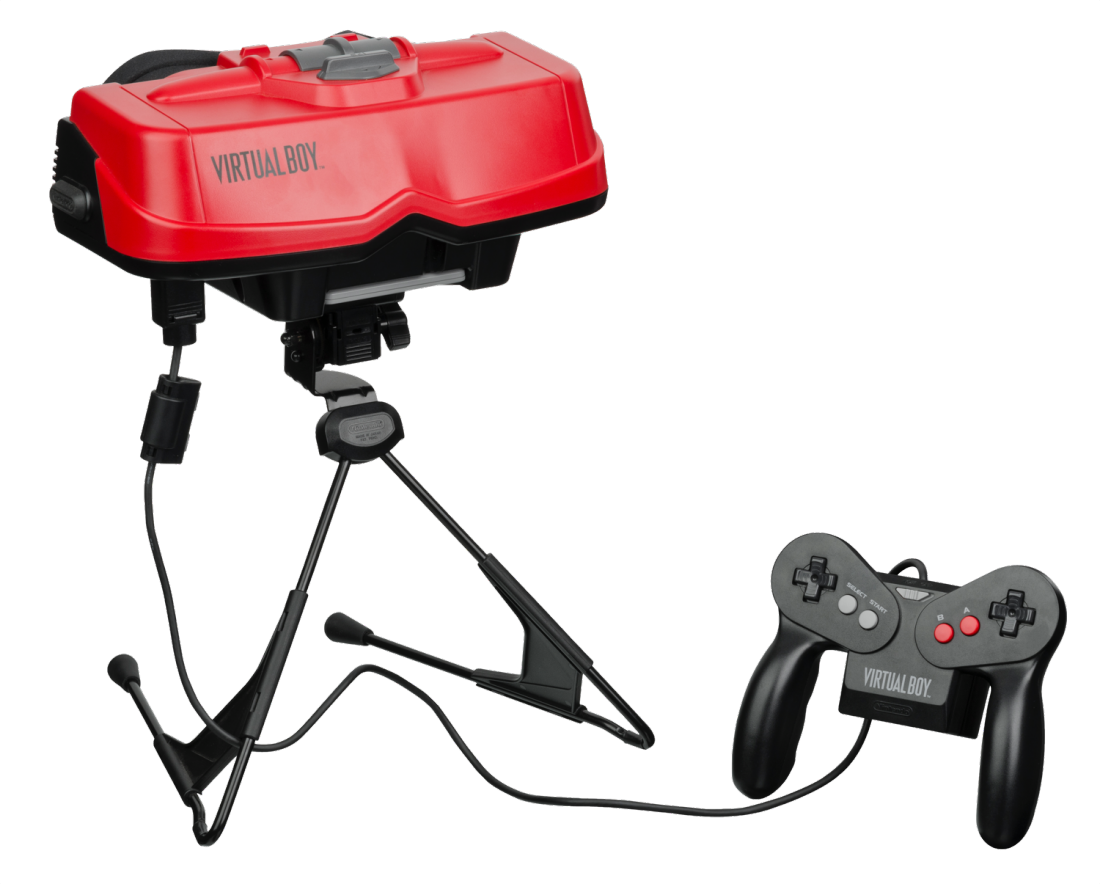}
    \end{minipage}
    }
    \subfigure[Sony Glasstron (1996)]{
    \begin{minipage}[htbp]{0.3\linewidth}
    \includegraphics[height=3cm]{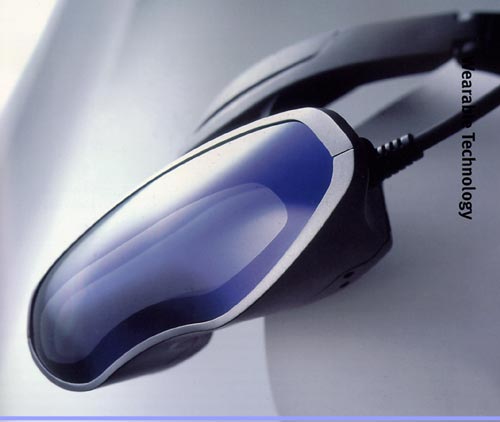}
    \end{minipage}
    }
    \caption{Popular HMDs in 1990s}
    \label{fig:hmd1}
    \end{figure*}

    Firstly proposed by Heilig and Sutherland in the 1960s~\cite{heilig1962sensorama}\cite{sutherland1968head}, the HMDs gradually gained popularity and became the representation of virtual reality over the next several decades. The screens of the HMDs are attached to the viewer's head and directly display the scene to his/her eyes. Furthermore, the HMDs also track the viewer's head movement and rotation then dynamically change the views on his left and right eye, as a result, they can provide the viewers with seamless motion parallax and binocular parallax. At the early age of the HMDs, the equipment was cumbersome, low-resolution, and very expensive. Then during the decades of development, the HMDs have experienced two periods of rapid growth. In the 1990s, some console producers such as Nintendo, Sony, and Sega published their own commercialized HMDs which first brought this type of 3D display into the public (Figure~\ref{fig:hmd1}). However, due to the restriction of the computer graphics and display resolution, these displays were low-fidelity which couldn't provide the viewers with a good enough experience. 
    
\begin{figure*}[ht]
    \subfigure[Oculus Rift DK1 (2013) ]{
    \begin{minipage}[htbp]{0.5\linewidth}
    \includegraphics[height=2.9cm]{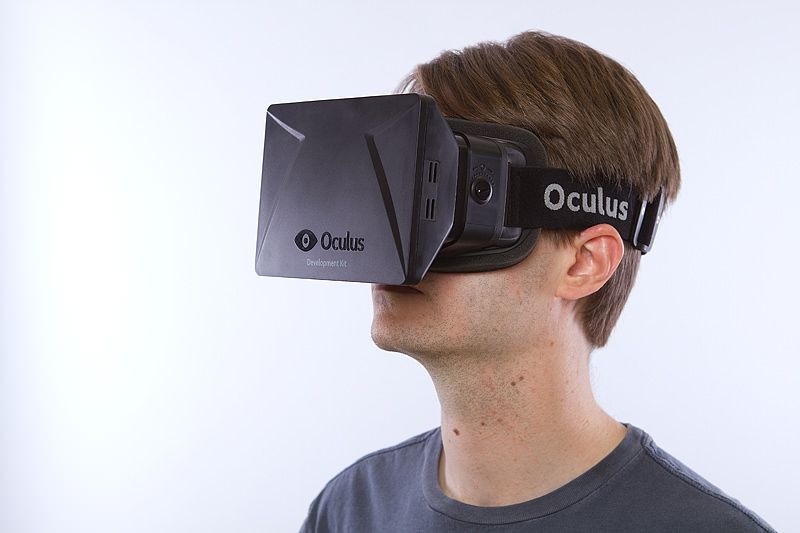}
    \label{fig:dk1}
    \end{minipage}
    }
    \subfigure[Microsoft Hololens (2016)]{
    \begin{minipage}[htbp]{0.45\linewidth}
    \includegraphics[height=3cm]{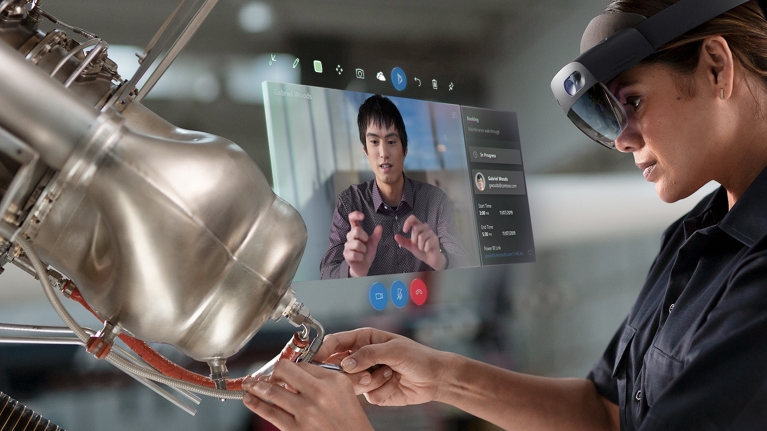}
    \label{fig:hololens}
    \end{minipage}
    }
    \caption{HMDs after 2013}
    
    \end{figure*}

    After the first boom, the industry hasn't published new HMDs for a long time until 2013. Over these years, the graphics, display, and video transmission techniques had incredible improvements. Based on these technical breakthroughs, another boom of the HMDs started. The first commercialized product in this stage was Oculus Rift DK1 (Figure~\ref{fig:dk1}). The headset had $1280\times800$ resolution, which is 100 times pixels more than the last generation HMDs. During the next few years, more companies published their HMDs with higher resolution, larger Field of View (FOV), and easier setup process. For example, the Pimax 8K X\footnote{\url{https://www.pimax.com}} HMD reached 8K ($3840\times2160\times2$) and 200-degree FOV; Google released their cardboard API\footnote{\url{https://arvr.google.com/cardboard/}} which made it possible to use some cardboard and a smartphone to make up a low-cost HMD; the standalone Oculus Quest 2 can even provide a similar experience as the desktop HMDs without connecting to any PC. The tracking system also evolved during this boom. Headsets are tracked accurately mainly through two types of approaches: station-based tracking system and inside-out tracking. The station-based tracking systems require tracking cameras (IR camera or webcam) mounted within a space to track the headset and controllers from 6 degrees of freedom (DOF). But due to the complicated setup procedure, these tracking systems are always used on dedicated desktop HMD systems. The inside-out tracking systems track users' movement through headset-attached webcams. By comparing the image difference between frames, the headsets can infer the user's location in the room. By now, this tracking method is more and more matured and applied to standalone HMDs. The Oculus Quest can even track user's hands and gestures accurately through the inside-out tracking system.
    
    Except for these Virtual Reality HMDs, another type of Mixed Reality (MR) HMDs such as Microsoft Hololens (Figure~\ref{fig:hololens}, Magic Leap\footnote{\url{https://www.magicleap.com/}}, and Microsoft Hololens 2\footnote{\url{https://www.microsoft.com/en-us/hololens}} came to the market. These MR HMDs not only provide the user with binocular and motion parallax but also allow viewers to see the real world through them (some researchers also named them as optical see-through HMDs\cite{milgram1994taxonomy}). 
    
    \subsubsection{Glass-required displays (GRDs):}
    Glass-required stereoscopic 3D displays are those displays that provide two views simultaneously and require the viewer to wear glasses for separating the views at their eyes. The light source of the display is normally placed at some distance to the viewers. According to the method of view separating, the eyewear could be classified as three types~\cite{urey2011state}:
    \begin{itemize}
        \item Color multiplexed eyewear: requires anaglyph glasses. The light source displays two views in blue and red respectively. By filtering out the views in a different color, each of the eyes can only see one view in the corresponding color on the glasses.
        
        \item Polarization multiplexed eyewear: requires polarized glasses. The directions of the two polarized glasses are orthogonal~\cite{urey2011state}. The light source renders one vertical view and horizontal view simultaneously. The view with different directions will be filtered out when the light goes through the polarized glasses.  
        
        \item Time multiplexed eyewear: the two eyes will never see two views simultaneously. The light source displays two views alternately at a specific frequency. The eyewear blocks the left and right eye at the same frequency so that at different times the eyewear allows one eye to see the image.
        
    \end{itemize}
    By utilizing these eyewears, the viewers can experience stereoscopic at a specific angle. Therefore, this technique was widely applied to 3DTVs and 3D cinemas to provide a large number of people with stereoscopic experience. However, if the GRDs only show 3D views to a single viewer, it's possible to recalculate the projection matrix to re-render the view based on the viewer's real-time position (known as off-axis view). 
    
\begin{figure*}[ht]
    \subfigure[CAVE immersive 3D display]{
    \begin{minipage}[htbp]{0.5\linewidth}
    \includegraphics[height=3cm]{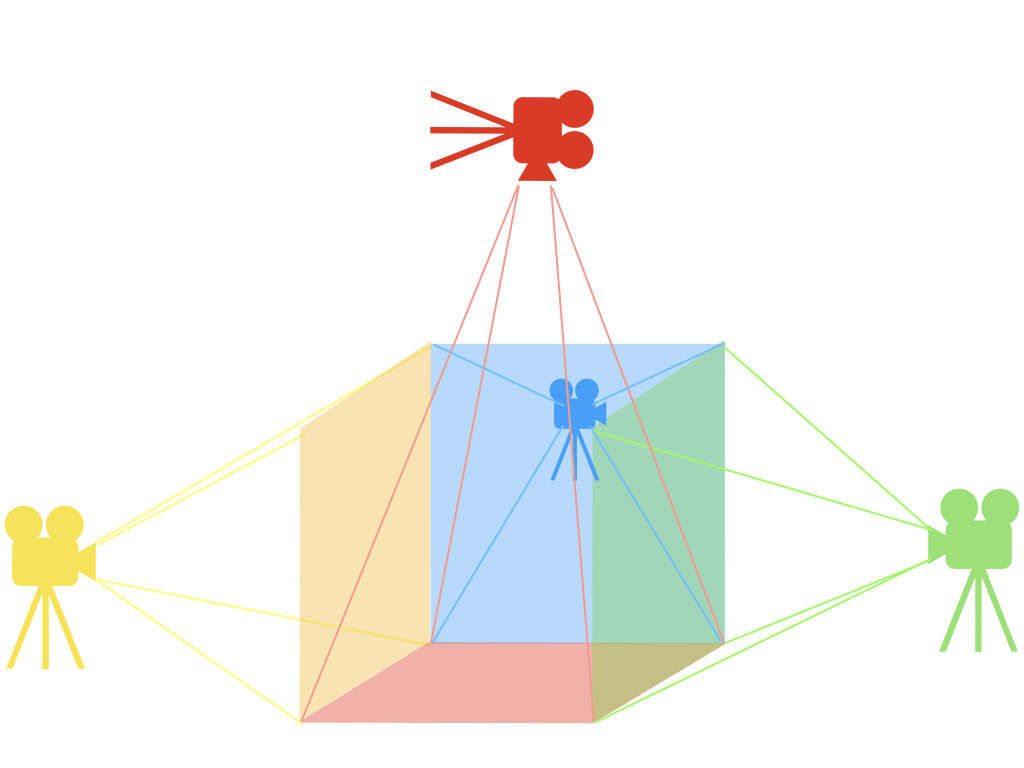}
    \label{fig:cave}
    \end{minipage}
    }
    \subfigure[Co-Globe 3D display\cite{zhou2018coglobe}]{
    \begin{minipage}[htbp]{0.5\linewidth}
    \includegraphics[height=3cm]{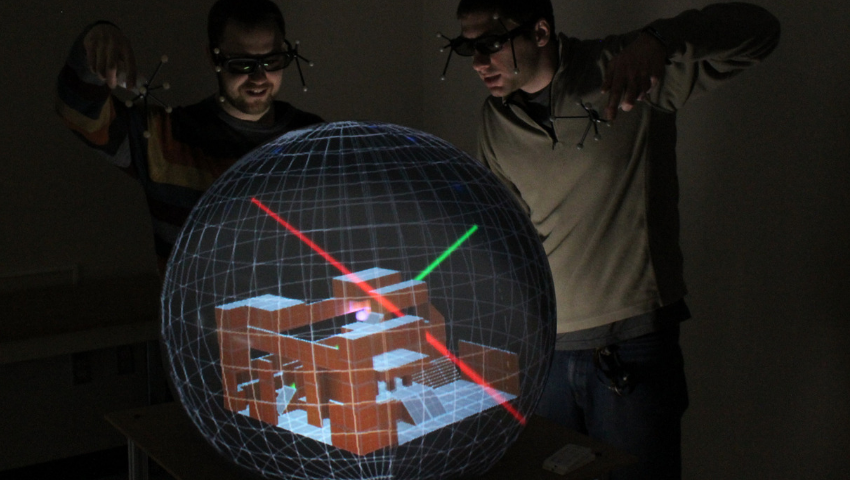}
    \label{fig:coglobe}
    \end{minipage}
    }
    \caption{GRDs system with motion parallax}
    
    \end{figure*}

    In 1992, Cruz et al. demonstrated their Cave Automatic Virtual Environment(CAVE)\cite{cruz1993surround} system. The CAVE system was a room-size immersive 3D GRD system that consisted of four sides of projection planes (Figure~\ref{fig:cave}). Each projection plane rendered two views alternately at 120hz and the viewer could see stereo views through a pair of shutter glasses. Furthermore, the system tracked the viewer's position through a tracker attached to the glasses. Thus the system was able to re-calculate the off-axis projection matrix and allowed motion parallax according to the viewer's position. 
    
    Except for the immersive 3D GRDs, some prior researches also brought motion parallax to small-scale GRDs through tracking the viewer's position. During the first boom of HMDs, Deering adapted a CRT monitor into a 3D display~\cite{deering1992high}. The system was originally designed to solve the low Pixel Per Degree (PPD) problem of the HMDs at that time. The system was equipped with time-multiplexed eyewear and a head tracker that allowed seamless motion parallax and binocular parallax. A similar system was proposed next year by Ware et al~\cite{ware1993fish}. Through the FTVR displays, viewers can move their heads within a restricted range while seeing a pair of off-axis stereo views. Therefore, Ware named this type of system as "Fish Tank VR" (FTVR) 3D displays. In 2018, an advanced adaption of the FTVR "Co-Globe"~\cite{zhou2018coglobe} was demonstrated on SIGGRAPH. The display had a spherical projection plane as a screen and had 360 degrees viewing angle with motion parallax and binocular parallax. Besides, Co-Globe also allowed two viewers to view and control it simultaneously. 
    
    \subsection{Monoscopic/non-stereoscopic 3D displays for motion parallax}
    \label{sec:mono}
    
    Some researches pointed out that motion parallax is more important than binocular parallax in 3D visualization~\cite{ware1993fish}\cite{sutherland1968head}, meanwhile, the invasive eyewears may bring the viewers with inconvenience in some situation; therefore, only providing motion parallax seems to be a good trade-off for some 3D display systems. The monoscopic or non-stereoscopic 3D displays are those 3D displays that provides only motion parallax but no binocular parallax. 

\begin{figure}[htbp]
    \centering
    \begin{minipage}[t]{0.48\textwidth}
    \centering
    \includegraphics[height=3cm]{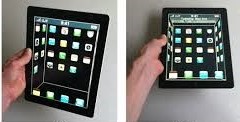}
    \caption{Mobile FTVR display\cite{francone2011using}} 
    \label{fig:mftvr}
    \end{minipage}
    \begin{minipage}[t]{0.45\textwidth}
    \centering
    \includegraphics[height=3cm]{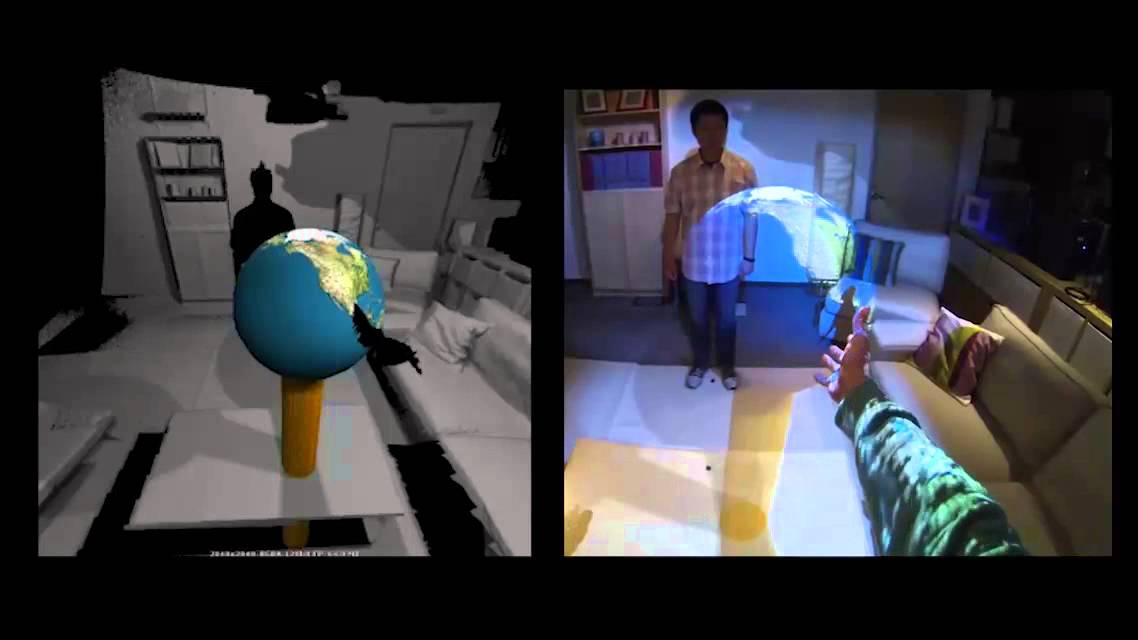}
    \caption{Microsoft SAR display system\cite{benko2014dyadic}}
    \label{fig:sar}
    \end{minipage}
    
    \end{figure}

    One type of monoscopic 3D display was the variation of the off-axis GRDs system. This type of 3D display system removed eyewears but enable the motion parallax by tracking the viewer's head position. Webcam-based face tracking was a widely applied solution on both desktop FTVR~\cite{rekimoto1995vision} and mobile FTVR~\cite{francone2011using} (Figure~\ref{fig:mftvr}). The webcam tracking was non-invasive but low-precision since the system was only able to track viewers from 2 degrees of freedom (DOF). To provide more accurate head tracking, Johnny Lee hacked the Nintendo Wii Remote controller and applied it as an infrared-red (IR) camera tracking system to the FTVR display~\cite{lee2008hacking}.
    After Microsoft released their IR camera based body-tracking camera Kinect, it was widely utilized as trackers for 3D displays. KAVE display system~\cite{gonccalves2018kave} was a monoscopic adaption of CAVE immersive 3D display. The head tracker of the CAVE display was replaced by Kinect body tracking so that the viewers were not required to wear any gears on their heads. 
    
    Another type of monoscopic 3D display aimed to create immersive Spatial Augmented Reality (SAR) environments in an enclosed space. The SAR systems exploit spatially-aligned optical elements such as projector-beam to overlay the augmented graphics onto the real environments without distracting the users from the perception of the real world~\cite{bimber2005spatial}. A good example of the SAR display was demonstrated by Microsoft in 2014~\cite{jones2014roomalive}\cite{benko2014dyadic} (Figure~\ref{fig:sar}). The system augmented a living room into an interactive environment by projectors. Multiple Kinect depth cameras were utilized to reconstruct the room and track the viewers. The virtual objects were projected directly onto the real objects according to the viewers' position. Furthermore, due to the room was reconstructed by the Kinects, the virtual objects could not only interact with the viewers but also with real objects in the room.

    \subsection{State of the Art:Next-generation Autostereoscopic displays and light field displays}
    \label{sec:LFD}
    Most of the 3D displays such as HMDs and GRDs require viewers to attach some forms of gears to their heads. These head-worn gears make it inconvenient due to the viewers have to put on and off the devices frequently. The autostereoscopic displays (ADs) aim to provide a glass-free 3D experience. An autostereoscopic display normally consists of two parts, a light source for rendering multiple views and a view splitter for separating the views. The light source of the ADs is similar to GRDs, which are usually projectors or flat-plane screens (LCD/LED/OLED panel). There are mainly three types of splitters: parallax barriers, lenticular lens, and retroreflectors. Different from the eyewears, these splitters separate the views at the displays rather than at the eyes. According to the number of views renders on the display, the ADs could be classified as two-view ADs, multi-view ADs, and super-multi-view ADs/light field displays~\cite{urey2011state}\cite{dodgson2005autostereoscopic}. 
    
    \subsubsection{Two-view autostereoscopic displays:}
    The two-view autostereoscopic displays are the earliest computer-generated autostereoscopic displays. They only display one pair of stereoviews. For most of the two-view ADs, the stereoviews are located at a steady position or repeat themselves within the viewing zone. One or multiple viewers can see the stereo views if they are in the correct viewing zone. But motion parallax is also important for a 3D display, therefore, different head-tracked two-view ADs were developed according to different light sources and splitters. Through tracking the viewer's head position, some displays dynamically turned on/off the barrier stripes to shift the viewing zones towards the viewers~\cite{perlin2000autostereoscopic}; some displays only utilized a movable stage which could shift the light source (projector) physically to create lateral motion parallax~\cite{surman2015head}. In the industry, Nintendo also produced their head-tracked two-view AD device "3DS" in 2011. These head-tracked two-view ADs allowed the viewers to move their heads within a restricted range while still seeing the 3D views. 
    
    \subsubsection{Multi-view autostereoscopic displays:}
    Multi-view autostereoscopic display systems produce multiple pairs of stereo views across the viewing zone. Viewer's left and right eye will see correct stereo views if they locate at different exit pupils. Furthermore, the multi-view ADs allow motion parallax within the viewing zones, but the number of views is too small to form seamless motion parallax~\cite{urey2011state}. The multi-view ADs are mostly adapted from the two-view ADs such as increasing the barriers interval~\cite{ando2005multi}, using multiple layers and utilizing time-multiplexed dynamic parallax barriers~\cite{choi2003multiple}. Comparing to the two-view ADs, the multi-view ADs enable multiple viewers to view the 3D contents from different aspects.
    
    \subsubsection{Super-multi-view autostereoscopic displays and light field displays:}
    Super-Multi-View(SMV) autostereoscopic displays are multi-view displays with a large number of views that produce continuous motion parallax. The viewing zone of the SMV displays normally covers a large angle which simulates a light field to some extents, therefore, some researchers also named the SMV autostereoscopic displays as light field displays (LFDs)~\cite{urey2011state}.
    The LFDs produce so many views simultaneously thus requires a very high-resolution light source to form clear images. Some projector-based LFDs are equipped with hundreds of projectors to ensure each of the views is high-quality and clear~\cite{lee2013optimal}. As off-the-shelf lenticular LFDs, the Looking Glass 15.6 inch~\footnote{\url{https://lookingglassfactory.com/product/15-6}} LFD also had an ultra-HD LCD panel as the light source. By now, the Looking Glass LFDs are the only off-the-shelf light-weight LFDs in the market.
     
\section{Comparative Analysis}
    \label{sec:discuss}
    With a full understanding of the current state of the art, a comparative analysis can be performed.
    In summary,  four aspects were chosen to measure the 3D displays:
    \begin{itemize}
        \item Depth cues: indicates which depth cues are provided by the 3D display. In this article, only binocular parallax and motion parallax will be discussed. 
        \item Immersion: indicates how much the viewers feel their presence in the virtual world~\cite{cruz1993surround}.
        \item Easy to set up: indicates if the system needs to be calibrated, mounted at a specific position, dedicated set up, or plug-in set up.
        \item Availability: indicates if the users need to do any action before they can see the 3D views.
    \end{itemize}
     
     \subsubsection{Depth cues:} The depth cues can assist viewers to understand the 3D scenes better so that more depth cues indicates a better 3D experience. According to Section~\ref{sec:dev}, the HMDs, some GRDs, most ADs, and LFDs which provide both binocular parallax and motion parallax provide the best 3D experience. 
     
    \subsubsection{Immersion:} The HMDs allow viewers to rotate their heads freely thus provide the highest immersion among the 3D displays. Some GRDs such as CAVE systems and 3D cinemas equip with large projection screens also provide an immersive experience. The monoscopic 3D displays and autostereoscopic displays normally have smaller screen size so that provide similar immersion experience to the viewers. High immersion displays are very suitable for displaying those 3D scenes that contain a number of details and don't require the viewers to focus on a specific point or area. The viewers should be able to choose what they want to see within the scene and explore it by themselves. By contrast, low-immersion displays are more suitable for displaying 3D scenes that need to be focus on so that viewers won't be distracted by the irrelevant information. 
      
    \subsubsection{Easy to set up:} The dedicated 3D display systems usually require a complicated setup and calibration process. These systems aren't normally supposed to be dismantled after setting up. Among these 3D displays, the GRDs are dedicated 3D displays thus they are suitable to be installed at labs, working places, or using for demonstration purposes on large events. Meanwhile, the HMDs also require dedicated space to have the best experience. And other 3D displays only require a plug-in setup just like the conventional 2D displays.
    
    \subsubsection{Availability:} To provide the highest availability, the 3D display must be non-invasive, in another word, there shouldn't be any gears mounted on the user's body. Among these 3D displays, only monoscopic displays and autostereoscopic displays can provide high availability. The GRDs and HMDs have low availability due to the head-mounted gears and headsets. Taking on and off the headset spend a longer time than any other displays. Furthermore, although ears, there still a lot of research reported that users the headsets are more and more lightweight in recent y will suffer neck aches and sicknesses if they wear the headset for a long time~\cite{ahern2020effectiveness}\cite{sharples2008virtual}. 
    
    \subsubsection{Discussion}
    
    Among the state-of-art 3D displays, the HMDs, head-tracked GRDs, and LFDs provide the best 3D experience. The HMDs and GRDs are more suitable for "non-focus" 3D experience while ADs are more suitable for "no-distraction" 3D experience. Considering the setup cost and availability, both GRDs and HMDs have high set up cost and low availability, but the HMDs have. Furthermore, as non-invasive displays, the ADs and some monoscopic displays do not need viewers to have close contact with them, which could be safely used during some pandemics such as COVID-19. To summarize, the HMDs are the best immersive 3D displays while the ADs provide a similar 3D experience with lower cost for setup and usage.
    
    Therefore, through this review, it appears that autostereoscopic display (especially light field display) is a quite pristine research area with the rich potential to explore in the future. Comparing to the HMDs, these displays still lack awareness of the context. The immersive HMDs have inherent context awareness, which gathers information from the environment and user such as the user's head position and rotation. Then they can utilize the information to showing the 3D views directly to the viewers' position. By contrast, the current LFDs don't perceive the context in any shape or form. Their seamless motion parallax comes from rendering a large number of stereoviews at the same time, which results in considerable waste of these invisible views. Meanwhile, technically, the views of an LFD are independent of each other~\cite{pan2020adaptive}. But the current LFDs render the same 3D contents on each of the views. This is another waste of the simultaneous views. Therefore, researches on combining context awareness to reduce this waste are needed to be carried out.  By considering context-awareness in relation to LFD's six future directions occur naturally. 


\section{Future directions}
\label{sec:future}
With the comparative analysis completed, the next step is to examine what future directions are made possible. Six potential directions will be proposed examining what can be achieved using current LFDs in combination with context awareness.

\subsection{Reducing graphics consumption}
\begin{figure}[htbp]
 \centering
 \includegraphics[width=0.7\linewidth]{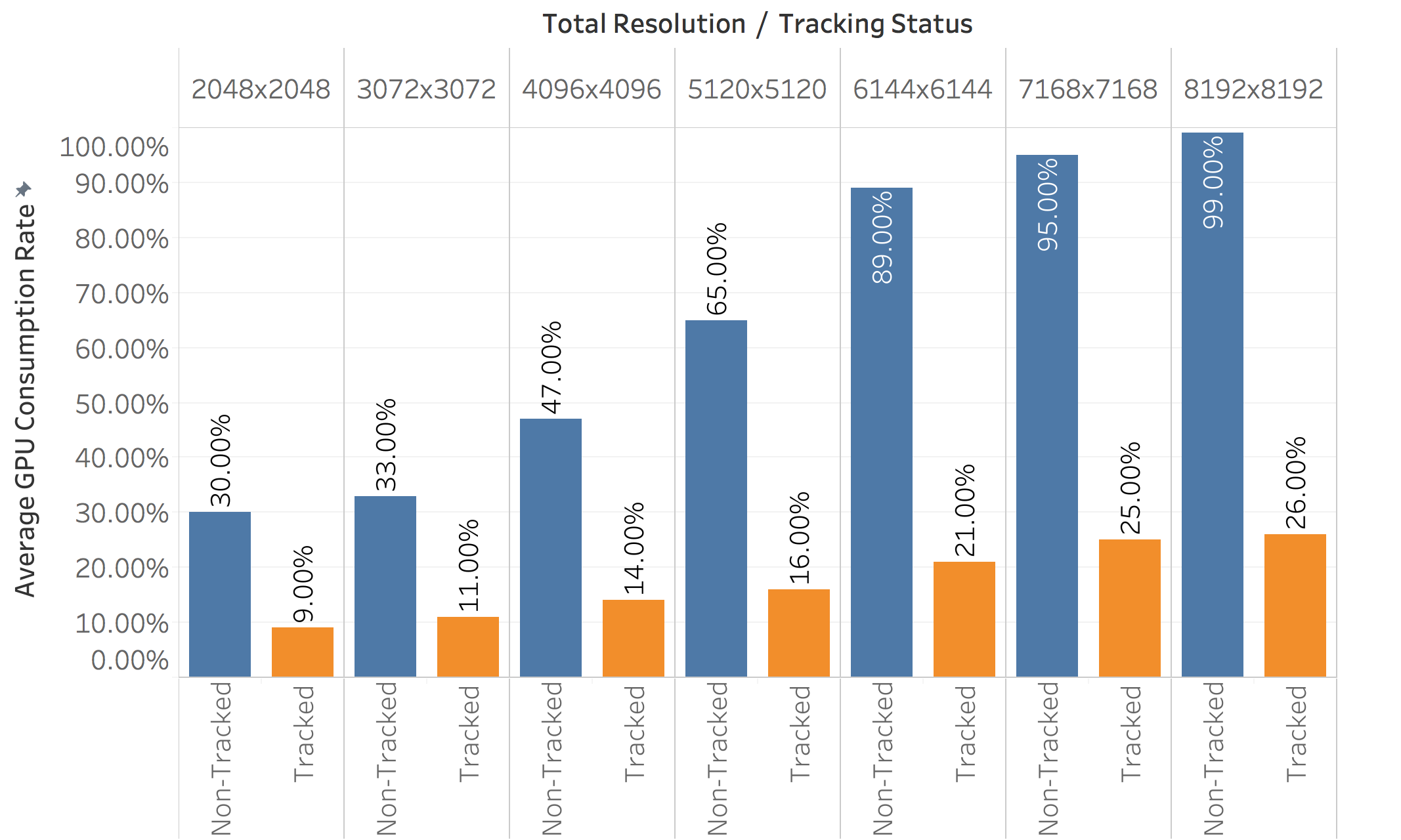}
    \caption{GPU Consumption}
    \label{fig:graph}
 \end{figure}

The current off-the-shelf LFDs render all the views simultaneously towards the viewing zones. However, the viewers normally won't be able to see all of the stereoviews at the same time so that considerable GPU resource is wasted. Prior work proposed an approach to save the wasted resource~\cite{pan2019exploration}. The approach was similar to a head-tracked two-view autostereoscopic display. The LFD system dynamically turns off the invisible views according to the viewer's real-time head position. Although the extra resource was used for head tracking, the GPU resource was significantly reduced (shown in Figure~\ref{fig:graph}). The work also simulated at different higher resolution, the result showed that the higher the future LFD, the more graphics consumption could be reduced. 

By combining with context awareness, more graphics consumption could be reduced. The future LFD systems should track viewers' gaze. For each viewing zone, the LFD will turn it off if there is no viewer looking at it. Furthermore, the future LFDs should also predict viewers moving tendency. The system could compare the viewer's position at a certain time interval to infer their moving direction and speed. Thus the next position that the viewer would appear at could be predicted. The routine prediction will allow the system to reduce the tracking refresh rate than less computational resources will be spent on head tracking. Therefore, the computational resources could be exploited to render larger and higher-resolution views. 

\subsection{Display Artifact Correction}
The conventional displays assume the viewers would look at them at specific viewing conditions such as fixed distance, viewing angle, and lighting. But in practice, these viewing conditions may vary according to different viewing scenarios. And the displays will have different effects under different viewing conditions. For example, the viewers would see a trapezoidal display rather than a rectangle-shape display when they stand on the sides of it (perspective offset); the display with the same brightness looks darker under the sunlight than it under the lamplight. The viewers had to adjust the displays manually to see clear and bright views. As context awareness and human-centered computing was introduced to mobile computing, people started to pay more attention to these demands, and different solutions were proposed to correct the artifacts (distortions). A method to dynamically control the brightness and color of the display was proposed~\cite{bell2016ambient}, The method suggested the displays to perceive the light condition by the webcam which didn't require any extra hardware support. Therefore, it was rapidly applied to the commercialized mobile displays during the past few years. The perspective offset correction method~\cite{bauer2016anatomical}, by contrast, required extra components such as the mechanical arms thus haven't been widely used for view correction. 

\begin{figure*}[ht]
    \subfigure[A 3D illusion street art\protect\footnotemark]{
    \begin{minipage}[htbp]{0.32\linewidth}
    \includegraphics[height=3cm]{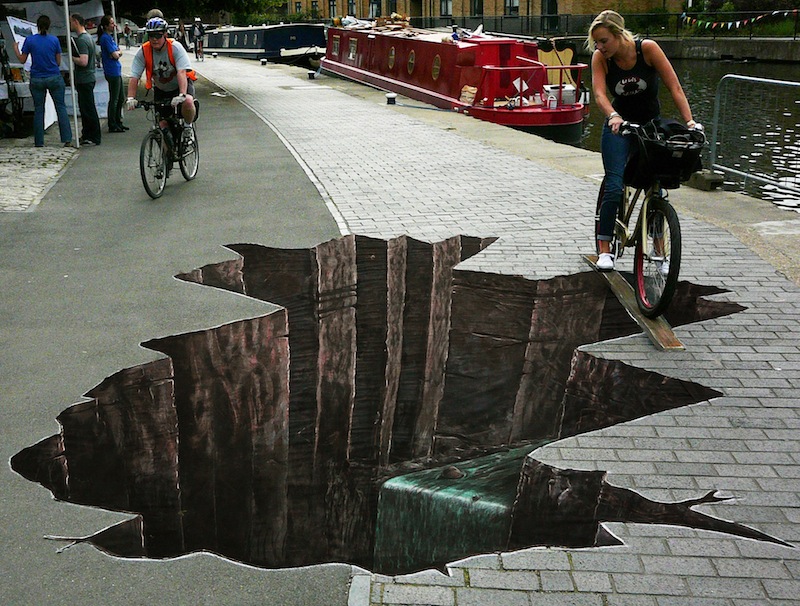}
    \label{fig:illusion}
    \end{minipage}
    }
    \subfigure[The front view]{
    \begin{minipage}[htbp]{0.30\linewidth}
    \includegraphics[height=3cm]{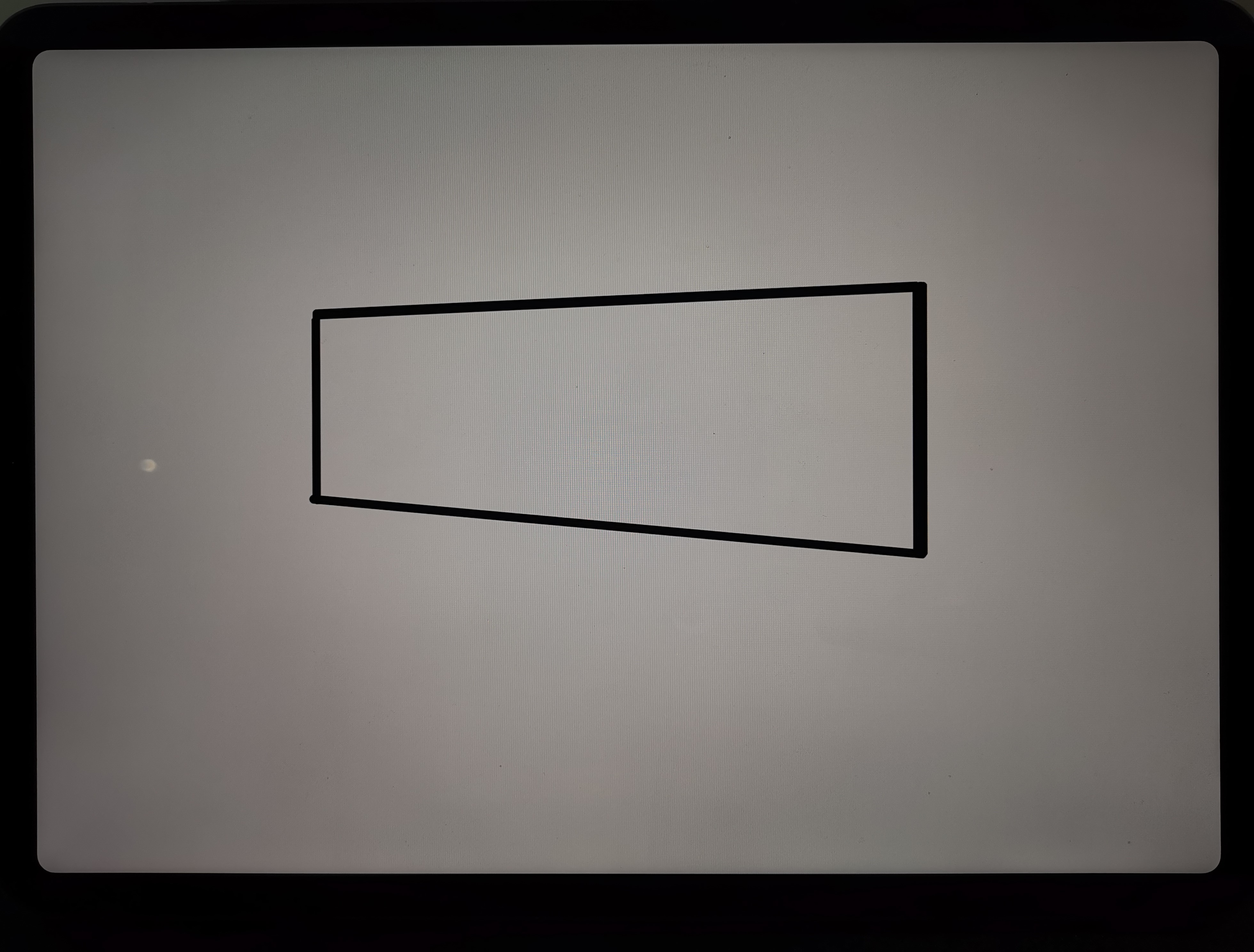}
    \label{fig:front}
    \end{minipage}
    }
    \subfigure[The side view]{
    \begin{minipage}[htbp]{0.22\linewidth}
    \includegraphics[height=3cm]{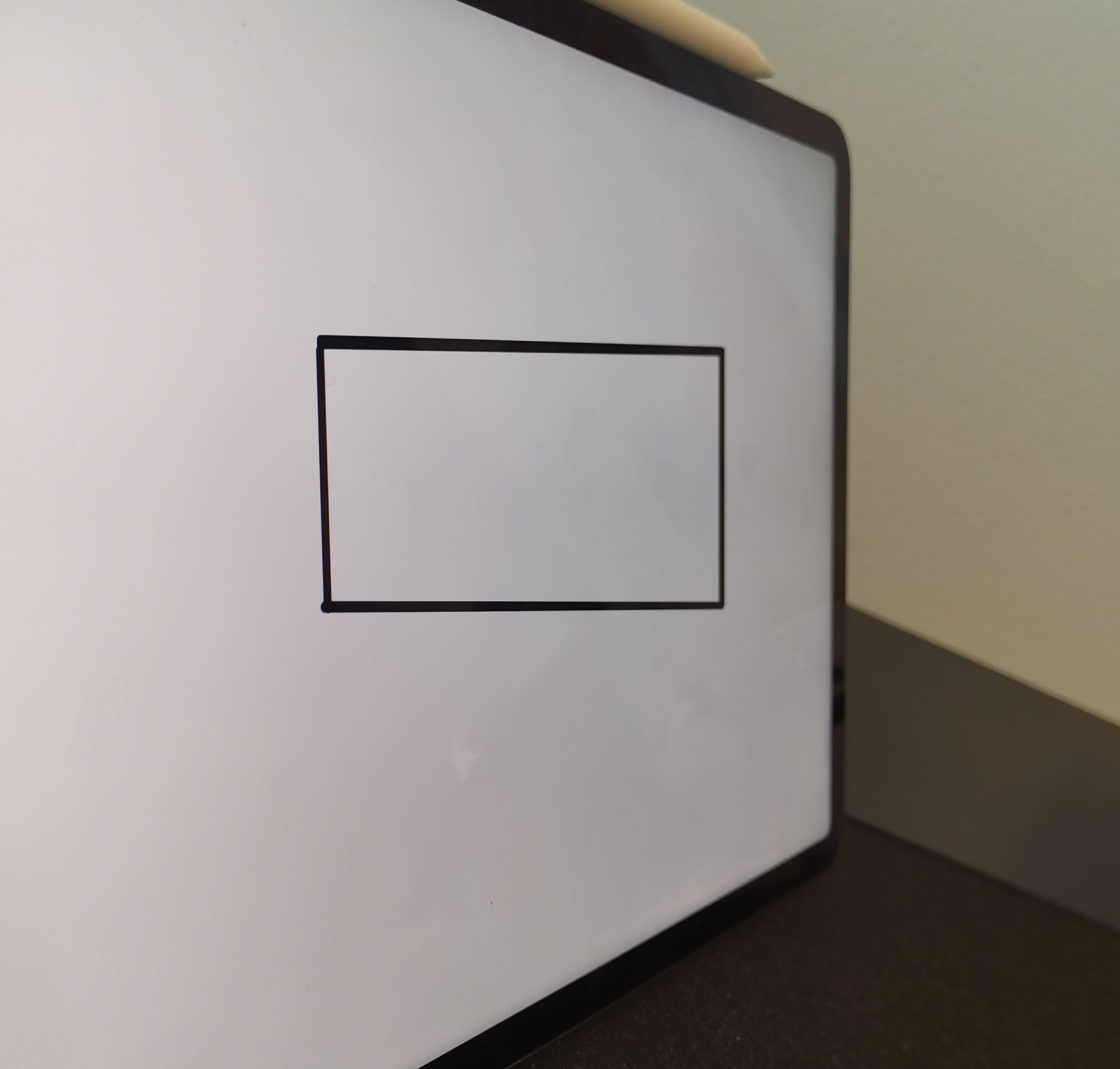}
    \label{fig:side}
    \end{minipage}
    }
    \caption{Principle of artifact correction}
    \label{fig:artifact}
    \end{figure*}
    \footnotetext{\url{https://www.opticalspy.com/street-art.html}}
    
Aiming for correcting the perspective offset artifact without requiring any external mechanical components, the correction method could be achieved purely at the software level in future light field displays. The principle of the approach is similar to the 3D illusion street arts (as Figure~\ref{fig:illusion} shown); the viewer will see the correct view from a specific perspective. This approach is to generate the illusion images according to the viewer's real-time eye position. Figure~\ref{fig:front} and Figure~\ref{fig:side} illustrated the principle of the approach on a 2D display. To show a rectangle for the viewing zone on the side of the display (Figure~\ref{fig:side}), the system should render a trapezoid shape (Figure~\ref{fig:front}) to counter-balance the distortion. By tracking the viewer's eye position and dynamically changing the rendered image on the LFDs, the viewer will be able to see a pair of corrected stereo views wherever his/her stands. In addition, the proposed approach will and only will work perfectly on the future LFDs. For a non-stereoscopic 2D display, it can only show one view at the same time, therefore, the correction can only be applied onto one of the eyes, which will still form an artifact overall.

\subsection{Privacy control}
\label{sec:privacy}
Modern people more and more rely on the computing systems in our daily life. These computing systems store a large quantity of our private information including our identity details, private messages, and intelligent properties (IPs). The previous researches have already proposed a lot of approaches to protect our data privacy from the information security point of view. But as the final media between computing systems and the users, the displays also play an important role to stop others from spying on our private data. The privacy control mechanism that is discussed within this paper is purely the display end. The conventional method to prevent others to view user's private data is to use security pins. In recent years, some biological information such as fingerprints and face IDs are also utilized to ensure user's privacy. However, the security check cannot stop the others from spying on the displays beside the user. User's private information could be gathered by a glimpse of an unauthorized person. Moreover, the displays won't be able to stop other spying user's data when the user forgets to lock the screen before he leaves. To avoid these privacy exposures, it is imagined a new privacy control method when people start to use LFDs in the future. The method is based on that different views on an LFD can show different 3D contents. The system will classify viewer's private data and the authorized viewers into different security levels in advance. The viewers are only allowed to view the information with the corresponding security level. During the runtime of the system, it will keep perceiving the surrounded environment and identify the viewers who are staring at the display. For those viewing zones being staring at, the display will filter out the information that doesn't match the viewer's identity. Therefore, the display will only show the private data to the owner instead of being seen by others. In particular, the display system should also temporarily block the information when someone else stands in the same viewing zone with the owner to prevent the information disclosure from behind. 

\subsection{Vision impairment correction}
\label{sec:disability}
The LFDs can also help people who have vision impairment (myopia, presbyopia, and irregular blurring in vision) to read properly without a need for glasses or contact lenses. With the implementations of 4D prefiltering algorithms, the LFD can process the target image to be desired vision-corrected, even for the eyes that are hard to be correct with eye-wears. \cite{huang2014eyeglasses} The algorithms enhanced LFDs can be applied on any displays, such as telestrator, laptop, or smartwatch, which allows user to perceive streams between multiple devices with bare eyes. The LFDs should be tailored for the consumer's eye condition and daily usage habit (watching direction, and the distance between the eye and the display), and the manual adjustment would cause extra workload to the user's daily usage. However, with nowadays face recognition and face tracking AI technique, the LFDs can calculate the distance of sight based on the face motion data and get the user's eye condition information by face recognition to update the pre-filter arguments in real-time, so that the device can be used by multiple users without trouble. For presbyopia patients, the LFDs may have the chance to let them get rid of eye-wear forever in the future, which may benefit the user in many daily life scenarios. Patients with vision impairment also have trouble when they are using the Virtual Reality or Augmented Reality HMD devices without their glasses or eye contact lenses. The application of LFDs in the Virtual Reality or Augmented Reality HMD devices can solve the convergence-accommodation conflict inherent issue to improve visual comfort with accurate reproduction of all depth cues~\cite{huang2015light}, and provides the possibility to increase the holographic rendering quality and account for optical defects between the light source and retina~\cite{maimone2017holographic}. LFD displays abilities to help with vision impairment may result in the future that these displays could be mandatory to fulfill regulatory requirements, analogous to mandatory accessibility ramps for modern buildings.

\subsection{Personalised Advertising}
\label{sec:ad}
The modern advertising system uses users' behavior data that is collected from their inputs (search term, clicking hot spot, purchasing record, etc.) Based on those anonymous data, the system can generate a profile of the user, and then, recommend the advertisement that meets the requirements to the profile of the user, which means, the user would have a higher possibility to click the advertisements.

Due to the special characteristic of the LFDs, the content rendered on the display can be different in view angles\cite{pan2020adaptive}. So is to say, multiple users can see customized images or videos on one LFD based on their positions. As mentioned in Section~\ref{sec:disability}, LFDs enhanced with facial recognition algorithms have the ability to track observers' identities and gather their information from the database. By adopting this technology, a giant advertising LFD may have the ability to identify multiple passersby and recommend specific contents to each of them based on their data, just like what has already been achieved on the internet. If so, one advertising board can serve massive users simultaneously with high-quality stereo content, which will not only increase the efficiency of advertising but also save space and energy of the public resources. A functional scenario might be like this: A advertising board was staying idle until the camera senses a passerby coming close. Then based on the user data, it starts rendering the target advertisement to the user only in one sight angle. Then more passerby coming close, it either renders a specific content for a user in one angle or renders some general backup contents if more than one users are standing too close to each other. 

However, a big argument and ethical problem are about whether a user is willing to be collected data for their personalized advertisement. Especially in the LFD's case, the facial recognition process may have the risk to expose the user's location, preference, or even personal data to the public environment, as the passerby who stands close enough will share the same view as the user does. To avoid privacy and ethical problem, the public advertising device should perform more generally. The facial recognition camera should only detect general biological data, such as height, gender, age-range, or gaze, from the users by machine learning. The collected data should not be stored on any platform in any shape or form.  

\subsection{Distributed Rendering}
\label{sec:distributed}

\begin{figure}[htbp]
 \centering
 \includegraphics[width=0.8\linewidth]{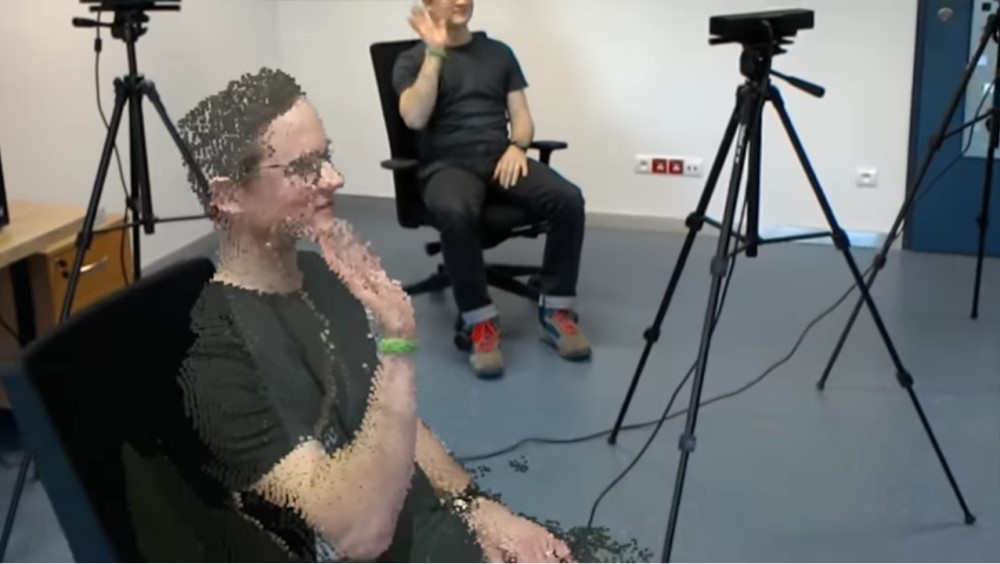}
    \caption{Point cloud telepresence on HoloLens\protect\footnotemark}
    \label{fig:pointcloud}
 \end{figure}

Because of the recent COVID-19 pandemic, it has become essential for people to communicate through an online chatting system. The world has witnessed an explosion in the use of online videoconferencing platforms such as ZOOM and Google Meet being more and more important for our social life. However, those systems have limitations in supporting nonverbal communication cues such as 3D spatial reasoning, and the movement of interlocutors~\cite{kim2012telehuman}. The solution is to track users' movement in real-time. One direction is to capture and transmit high-quality 3D point clouds using IR cameras~\cite{kowalski2015livescan3d}. Researchers have found several ways to present this form of human telepresence, including a cylindrical Display by Kim et al., and Hololens~\cite{hololens_youtube}. Human telepresence shows a great potential to benefit future human communication in many aspects, including teleteaching and teleconferencing, even under the huge pressure of the pandemic situation. However, the cylindrical Display cannot provide a stereoscopic view for the observer as it is a curved 2D projection. The usage of Hololens can perfectly solve the problem, showing the streaming 3D cloud point in a stereoscopic view(Figure~\ref{fig:pointcloud}). But for educational purposes, it is expensive to apply the Hololens to class-scale multiple users, and the HMDs have to be reused between users, which increases the potential of the virus spreading during a hard time. 

\footnotetext{\url{https://www.youtube.com/watch?v=Wc5z9OWFTTU}}

Compared with those display solutions, the LFDs have advantages like contact-free, providing stereoscopic view without additional gear, and multiple usages at the same time. Cserkaszky et al. \cite{cserkaszky2018real} implemented a real-time light-field 3D telepresence system with a couture system assembled by 96 cameras and a human-size display system. The system provided a high level of the sense of presence by smooth horizontal motion parallax and human-size scale display through the Local Area Network (LAN). However, this system had several limitations: it replied on a high-bandwidth LAN, which limits the actual remote scenario; the stream frame rate had been limited to 25 due to the capacity of the camera broadcaster; the camera system could not be moved. The usage of 3D point clouds or a virtual avatar can solve the above problem as the volume of the transferred data is much lower and the depth cameras are mostly portable nowadays. Although it is hard to create one big LFD with a single LED panel, more than one small LFDs can be assembled to achieve a similar visual effect. The point cloud data or the facial capture data can be used to render the human telepresence distributively through each display, which will not only separate the rendering pressure to multiple devices but also make the system scalable and customizable. 

Another direction to track users' movement in real-time to achieve telepresence is using a virtual avatar. The virtual avatar model can be stored locally, so, the data that needs to be transferred through the internet will only be the facial transform information. Compared with the 3D point cloud, this method will save massive internet bandwidth. As for the model details, the photogrammetry technique can scan a human face and create a high fidelity model, and the modeling tools nowadays, such as Maya and Blender, can create face joints and blend shapes, used for facial motion capture, automatically supported by AI technique. Except this, a high fidelity human model can be generated easily with next-gen software like Metahuman\footnote{\url{https://www.unrealengine.com/en-US/digital-humans}}. Moreover, the facial motion capture can be achieved by one depth camera, while the 3D point cloud needs 2 and more to achieve the same level of model fidelity as one depth camera cannot capture the details on the backside. The body motion capture can either be achieved by a single depth camera, mocap suit, or a Virtual Reality system.

\section{Conclusion}
\label{sec:conc} 
The work presented in this paper explores the future directions of non-HMD-based 3D displays. First, the paper generally discusses the current state of art in 3D displays. Once this background was established, it was possible to conduct a comparative analysis to explore what potential future directions could gleam from the Current State of the Art. These future directions could be applied specifically to Light field displays as they appear to be an area with the least research conducted on. 

This paper proposes a number of possible future research directions for LFDs. One drastic issue that stops the potential of light field displays dead in their tracks is the huge computational requirements of support 10's or more possible view frustums. Through context-awareness only the visible viewpoints are rendered to reduce graphics consumption and make this technology sustainable at high resolutions. New research can explore how to correct for display artifacts caused by a user either viewing the display at an incorrect angle or even in incorrect lighting conditions. Displays could in the future only display information to one user even when another user is sitting right beside them using privacy control. By knowing which user is looking at a display, the screen could automatically correct their vision so they do not need to wear glasses. The dream of personalized outdoor advertising (not unlike our advertising experiences online), where multiple people could walk by the same shop display but based on machine learning, the advertisement to each of them would be different even if they stood right next to each other. This of course has ethical implications because not only would the advertising be personalized but from the passerby's perspective each of their objective realities would be different.

Finally, with full context awareness and the capability to distribute rendering across multiple displays and devices, LFD could achieve the goal of truly allowing real telepresence experiences with full-size participants. This work outlined in the future directions is just a small step forward in fulfilling Ivan Sutherland's dream of the Ultimate Display  
\textit{``with appropriate programming, such a display could literally be the Wonderland into which Alice walked"}\cite{sutherland1965ultimate}, but this paper demonstrates that devices are now ready and it is simply the programming that is lacking; let's fix that.

%
%

\bibliographystyle{unsrt}

\end{document}